\documentclass[twocolumn,nofootinbib,longbibliography]{revtex4-1}

\usepackage[english]{babel}
\usepackage{appendix}
\usepackage{natbib}

\usepackage{amsmath}
\usepackage{mathrsfs}
\usepackage{amssymb}
\usepackage{amsfonts}
\usepackage{dsfont}
\usepackage{IEEEtrantools}
\usepackage{bm}
\usepackage{amsthm}

\usepackage[breaklinks=true]{hyperref}
\usepackage[inline]{enumitem}

\makeatletter
\usepackage[threshold=1, autopunct, autostyle]{csquotes}
\@ifclassloaded{beamer}
  {}
  {
    \bibpunct{[}{]}{,}{n}{}{,}

  }
\makeatother

\usepackage{pgfplots}
\usepackage{tikz}
\usetikzlibrary{shapes,calc,positioning,arrows.meta}

\pgfdeclarelayer{bg}    \pgfsetlayers{bg,main}

\tikzset{colorbox/.style={thick, rounded corners=2pt, text height=1.7ex,text depth=.25ex, draw=#1!70!black, fill=#1!30}}
\tikzset{colorboxS/.style={thick, rounded corners=2pt, text height=1.2ex,text depth=.25ex, draw=#1!70!black, fill=#1!30, font=\footnotesize}}
\tikzset{colorboxXS/.style={thick, rounded corners=2pt, text height=0.7ex,text depth=.10ex, draw=#1!70!black, fill=#1!30, font=\tiny}}
\tikzset{roundedbox/.style={thick, rounded corners=2pt, text height=1.7ex,text depth=.25ex, draw=black}}
\tikzset{CBedgy/.style={thick, text height=1.7ex,text depth=.25ex, draw=#1!70!black, fill=#1!30, font=\ttfamily}}
\tikzset{colorelement/.style={thick, rounded corners=2pt, draw=#1!70!black, fill=#1!30}}
\tikzset{quote/.style={thick, draw=black!70, rounded corners=2pt, font=\footnotesize, text width=#1, text opacity=1, opacity=.8, fill=white},
  quote/.default=4.5cm}
\tikzset{quoteS/.style={thick, draw=black!70, rounded corners=2pt, font=\scriptsize, text width=#1, text opacity=1, opacity=.8, fill=white},
  quoteS/.default=4.5cm}
\tikzset{quoteXS/.style={thick, draw=black!70, rounded corners=2pt, font=\tiny, text width=#1, text opacity=1, opacity=.8, fill=white},
  quoteXS/.default=4.5cm}
\tikzset{quoteNB/.style={font=\footnotesize, text width=#1, text opacity=1, opacity=.8, fill=white},
  quoteNB/.default=4.5cm}
\tikzset{comment/.style={thick, draw=black!70, rounded corners=2pt, font=\scriptsize\itshape, text width=#1, text opacity=1, opacity=.8, fill=white},
  comment/.default=4.5cm}
\tikzset{commentS/.style={thick, draw=black!70, rounded corners=2pt, font=\tiny\itshape, text width=#1, text opacity=1, opacity=.8, fill=white},
  commentS/.default=4.5cm}
\tikzset{commentVW/.style={thick, draw=black!70, rounded corners=2pt, font=\scriptsize\itshape, text opacity=1, opacity=.8, fill=white}}
\tikzset{commentSVW/.style={thick, text height=0.7ex,text depth=.10ex, draw=black!70, rounded corners=2pt, font=\tiny\itshape, text opacity=1, opacity=.8, fill=white}}
\tikzset{conn/.style={thick, shorten <=#1, shorten >=#1}}
\tikzset{tconn/.style={shorten <=#1, shorten >=#1}}
\tikzset{gconn/.style={thick, shorten <=#1, shorten >=#1, draw=gray!80}}
\tikzset{arr_node/.style={pos=0.5,above,font=\scriptsize, sloped}}
\tikzset{ctag/.style={thick, dashed, rounded corners=2pt, text height=1.7ex,text depth=.25ex, draw=#1!70!black, fill=#1!30, font=\ttfamily}}

\tikzset{ist/.style={thick, shorten <=4pt, shorten >=4pt, arrows = {-Bracket[reversed,round]}}}
\tikzset{istgleich/.style={thick, shorten <=4pt, shorten >=4pt, arrows = {Bracket[reversed,round]-Bracket[reversed,round]}}}
\tikzset{nicht/.style={thick, dashed, shorten <=4pt, shorten >=4pt, arrows = {Bracket[round]-}}}

\definecolor{yorange}{HTML}{ff8c00}

\newcommand\blArrow{}
\def\blArrow[#1](#2);
  { \draw[#1] (#2) -- ++(.9,0) -- ++(-.9,-5) -- cycle; }
\newcommand\stdBlArrow{}
\def\stdBlArrow(#1)
  { \blArrow[line width=2pt, draw=red!70!black, fill=red!40, rounded corners=1pt](#1) }

\newcommand\redArrow{}
\def\redArrow(#1);
  { \draw[thick] ($(#1)+(-.3,.1)$) -- ($(#1)+(0,-.1)$) -- ($(#1)+(.3,.1)$);}
\newcommand\fSTS{}
\def\fSTS(#1);
  { \draw[fill=red!40, thick, draw=red!60!black] (#1) circle (.5);
    \draw[->] ($(#1)+(-.05,-.05)$) to ++(0,.4);
    \draw[->] ($(#1)+(-.05,-.05)$) to ++(.4,0);
    \draw[->] ($(#1)+(-.05,-.05)$) to ++(-.2,-.2);}
\newcommand\reduction{}
\def\reduction(#1);
  { \draw[fill=cyan!40, thick, draw=cyan!60!black] (#1) circle (.5);
    \draw[fill=black] ($(#1) + (-.2, .25)$) to ++(0.4,0) to ++(-.3,-.6) to cycle;}
\newcommand\customLegend{}
\def\customLegend(#1);
  {\fSTS(#1);
   \node[right of=#1] {fixed space-time structure};
   \draw[rounded corners=2pt] ($(#1)+(-.5,-.5)$) rectangle ($(#1)+(4, -4)$);}

\makeatletter
\newcommand*{\blitzset}{\pgfqkeys{/blitz}}\blitzset{
  height/.initial=4cm,
  width/.initial=2cm,
  breadth/.initial=.3cm,
  ratio/.initial=.4,
}

\pgfdeclareshape{blitz}
{
  \savedanchor\centerpoint{
    \pgf@x = .5\wd\pgfnodeparttextbox
    \pgf@y = .5\ht\pgfnodeparttextbox
  }
  \anchor{center}{\centerpoint}
  \anchor{north}{\centerpoint\pgfmathsetlength\pgf@ya{\pgfkeysvalueof{/blitz/ratio}*\pgfkeysvalueof{/blitz/height}}\advance \pgf@y by \pgf@ya}
  \anchor{south}{\centerpoint\pgfmathsetlength\pgf@ya{-(1-\pgfkeysvalueof{/blitz/ratio})*\pgfkeysvalueof{/blitz/height}}\advance \pgf@y by \pgf@ya}
  \anchor{east}{\centerpoint\pgfmathsetlength\pgf@ya{\pgfkeysvalueof{/blitz/breadth}}\advance \pgf@y by \pgf@ya\pgfmathsetmacro{\alpha}{atan(2*\pgfkeysvalueof{/blitz/ratio}*\pgfkeysvalueof{/blitz/height}/\pgfkeysvalueof{/blitz/width})}\pgfmathsetlength\pgf@xa{.5*\pgfkeysvalueof{/blitz/width}+\pgfkeysvalueof{/blitz/breadth}/tan(\alpha/2)}\advance\pgf@x by \pgf@xa}
  \anchor{west}{\centerpoint\pgfmathsetlength\pgf@ya{\pgfkeysvalueof{/blitz/breadth}}\advance \pgf@y by -\pgf@ya\pgfmathsetmacro{\alpha}{atan(2*\pgfkeysvalueof{/blitz/ratio}*\pgfkeysvalueof{/blitz/height}/\pgfkeysvalueof{/blitz/width})}\pgfmathsetlength\pgf@xa{.5*\pgfkeysvalueof{/blitz/width}+\pgfkeysvalueof{/blitz/breadth}/tan(\alpha/2)}\advance\pgf@x by -\pgf@xa}

  \backgroundpath{
    \centerpoint \pgf@xa=\pgf@x \pgf@ya=\pgf@y
    \pgfmathsetmacro{\alpha}{atan(2*\pgfkeysvalueof{/blitz/ratio}*\pgfkeysvalueof{/blitz/height}/\pgfkeysvalueof{/blitz/width})}
    \pgfmathsetlength\pgf@xb{.5*\pgfkeysvalueof{/blitz/width}-\pgfkeysvalueof{/blitz/breadth}/tan(\alpha/2)}
    \pgfmathsetlength\pgf@yb{\pgfkeysvalueof{/blitz/breadth}}
    \advance \pgf@xa by \pgf@xb
    \advance \pgf@ya by -\pgf@yb
    \pgfpathmoveto{\pgfpoint{\pgf@xa}{\pgf@ya}}
    \pgfmathsetlength\pgf@xb{\pgfkeysvalueof{/blitz/width}}
    \advance \pgf@xa by -\pgf@xb
    \pgfpathlineto{\pgfpoint{\pgf@xa}{\pgf@ya}}
    \pgfmathsetlength\pgf@yb{\pgfkeysvalueof{/blitz/breadth}+\pgfkeysvalueof{/blitz/ratio}*\pgfkeysvalueof{/blitz/height}}
    \pgfmathsetlength\pgf@xb{\pgf@yb*sin(90-\alpha)}
    \advance \pgf@xa by \pgf@xb
    \advance \pgf@ya by \pgf@yb
    \pgfpathlineto{\pgfpoint{\pgf@xa}{\pgf@ya}}
    \pgfmathsetlength\pgf@xb{2*\pgfkeysvalueof{/blitz/breadth}/cos(90-\alpha)}
    \advance \pgf@xa by \pgf@xb
    \pgfpathlineto{\pgfpoint{\pgf@xa}{\pgf@ya}}
    \pgfmathsetlength\pgf@yb{\pgfkeysvalueof{/blitz/ratio}*\pgfkeysvalueof{/blitz/height}-\pgfkeysvalueof{/blitz/breadth}}
    \pgfmathsetlength\pgf@xb{\pgf@yb*sin(90-\alpha)}
    \advance \pgf@xa by -\pgf@xb
    \advance \pgf@ya by -\pgf@yb
    \pgfpathlineto{\pgfpoint{\pgf@xa}{\pgf@ya}}
    \pgfmathsetlength\pgf@xb{\pgfkeysvalueof{/blitz/width}}
    \advance \pgf@xa by \pgf@xb
    \pgfpathlineto{\pgfpoint{\pgf@xa}{\pgf@ya}}
    \pgfmathsetlength\pgf@yb{(1-\pgfkeysvalueof{/blitz/ratio})*\pgfkeysvalueof{/blitz/height}+\pgfkeysvalueof{/blitz/breadth}}
    \pgfmathsetlength\pgf@xb{.5*\pgfkeysvalueof{/blitz/width}+\pgfkeysvalueof{/blitz/breadth}/tan(\alpha/2)}
    \advance \pgf@xa by -\pgf@xb
    \advance \pgf@ya by -\pgf@yb
    \pgfpathlineto{\pgfpoint{\pgf@xa}{\pgf@ya}}
    \pgfpathclose
  }
}

\newcommand*{\srefset}{\pgfqkeys{/sref}}\srefset{
  height/.initial=.5cm,
  width/.initial=.5cm,
}

\pgfdeclareshape{sref}
{
  \savedanchor\centerpoint{
    \pgf@x = .5\wd\pgfnodeparttextbox
    \pgf@y = .5\ht\pgfnodeparttextbox
  }
  \anchor{center}{\centerpoint}
  \anchor{north}{\centerpoint\pgfmathsetlength\pgf@ya{.5*\pgfkeysvalueof{/sref/height}}\advance \pgf@y by \pgf@ya}
  \anchor{south}{\centerpoint\pgfmathsetlength\pgf@ya{-.5*\pgfkeysvalueof{/sref/height}}\advance \pgf@y by \pgf@ya}
  \anchor{east}{\centerpoint\pgfmathsetlength\pgf@xa{.5*\pgfkeysvalueof{/sref/width}}\advance \pgf@x by \pgf@xa}
  \anchor{west}{\centerpoint\pgfmathsetlength\pgf@xa{-.5*\pgfkeysvalueof{/sref/width}}\advance \pgf@x by \pgf@xa}

  \backgroundpath{
    \centerpoint \pgf@xa=\pgf@x \pgf@ya=\pgf@y
    \pgfmathsetlength\pgf@xb{.5*\pgfkeysvalueof{/sref/width}}
    \pgfmathsetlength\pgf@yb{.5*\pgfkeysvalueof{/sref/height}}
    \pgfmathsetlength\pgf@yc{.75*\pgfkeysvalueof{/sref/height}}
    \advance \pgf@xa by \pgf@xb
    \pgfpathmoveto{\pgfpointpolar{20}{\pgf@xa}}
    \pgfpatharc{20}{170}{\pgf@xb}
    \pgfpathlineto{\pgfpointpolar{140}{\pgf@yc}}
    \pgfpathmoveto{\pgfpointpolar{200}{\pgf@xa}}
    \pgfpatharc{200}{350}{\pgf@xb}
    \pgfpathlineto{\pgfpointpolar{320}{\pgf@yc}}
  }
}
\makeatother

\makeatletter
\@ifclassloaded{beamer}
  {}
  {

    \theoremstyle{remark}
    
    \theoremstyle{definition}

  }
\makeatother

\newif\ifradical
\radicaltrue
\newif\ifbeta
\betatrue
\newif\ifuncertain
\uncertaintrue
\newif\ifcomments
\commentstrue

\newcounter{CtrSprachspiel}

\makeatletter
\def\p@paragraph{\thesection\,\thesubsection.}
\makeatother

\let\oldparagraph=\paragraph
\renewcommand\paragraph[1]{\oldparagraph{#1.}}

\makeatletter\begin{document}

\title{Scientistic Answers and Philosophical Questions}
\title{\Huge\bf\sf sex, lies,\\ and wigner's tape}
\title{\small --- for internal use only --- \\ \Huge\bf\sf sex, lies,\\ and wigner's tape}
\title{The Quest for Absolute Truth}
\title{No Truth Out There?}
\title{Truth Out There?}
\title{Ironic Science}
\title{About Truth and Lie in the Scientific Sense}
\title{Truth is made}
\title{Fabricated Reality}
\title{Everything you always wanted to know about Wigner's friend}
\title{No Observation-In-Itself}
\title{After the Measurement Problem is \\ Before the Measurement Problem}
\title{The Sensation of a Problem}
\title{\sc No master (key)\\ No (measurement) problem}

\author{Arne Hansen and Stefan Wolf}
\affiliation{Facolt\`a di Informatica, 
Universit\`a della Svizzera italiana, Via G. Buffi 13, 6900 Lugano, Switzerland}

\date{\today}

\begin{abstract}
\noindent
Can normal science---in the Kuhnian sense---add something substantial to the discussion about the measurement problem?
Does an extended Wigner's-friend \emph{Gedankenexperiment} illustrate new issues? 
Or a new quality of known issues?
Are we led to new interpretations, new perspectives, or do we iterate the previously known?
The recent debate does, as we argue, neither constitute a turning point in the discussion about the measurement problem nor fundamentally challenge the legitimacy of quantum mechanics.
Instead, the measurement problem asks for a reflection on fundamental paradigms of doing physics.

 \end{abstract}

\maketitle

\section{The (extended) measurement problem}\label{sec:introduction}
\noindent
Classical mechanics allows for a transparent picture of the world:
If the physical states were its furniture, then the \emph{thing-in-itself} would be immediately accessible:
Momentum and position of massive objects were, in principle, \emph{not} barred from immediate and simultaneous perceptual access.
Physics provided a powerful language: There were seemingly no boundaries to what it could describe---and with that: \emph{tame}.
Since the wake of quantum mechanics, the idea of an immediately accessible underlying reality described by physical theories has gotten cracks.
One of the problems besides non-locality~\cite{Bell1964} and contextuality~\cite{KS67}:
The quantum-mechanical formalism cannot simply be extended to an exhaustive description of the observer---if it were extended, then what would we make of an observer in a superposition state?

The problem is commonly exemplified with the \emph{Wigner's-friend experiment}~\cite{Wigner1961,deutsch1985quantum}: Wigner performs a measurement on a joint isolated system~$S\otimes F$ containing his friend~$F$ who, in turn, measures a system emitted by a source~$S$.
If the measurement bases of Wigner and his friend are chosen appropriately with respect to the emitted source state, then the probability distribution of Wigner's measurement of~$S\otimes F$ is ambiguous:
It depends on whether the friend's measurement is considered to be a quantum-mechanical evolution with a corresponding unitary operator on the joint system or it is taken to induce a ``collapse'' associated with a definite measurement result.

\emph{But what is the actual problem here?} 
Is there a contradiction at the heart of the quantum-mechanical description?
Is quantum mechanics in this sense fundamentally broken?
Is something wrong with quantum mechanics rather than with the expectation of a physical theory describing the real thing-in-itself?
The Frauchiger/Renner article~\cite{FR18} seems to provide an affirmative answer: 
The measurement problem is not just a peculiarity of quantum mechanics being probabilistic; 
the problem can be exemplified with a \emph{single-shot Gedankenexperiment}---in an \emph{extended Wigner's-friend experiment}.
The authors concluded in an earlier version that, therefore, ``single-world interpretations are inconsistent''~\cite{FrRen,BHW16}.
The conclusion relies on carrying over the inconsistency of a \emph{formalism} to \emph{one possible interpretation}~\cite{BW17}.

Just as conclusions from an extended Wigner's-friend experiment to interpretations are in doubt, so are claims that, because of the \emph{Gedankenexperiment}, quantum mechanics is more troubled than it has been before.
If quantum mechanics is regarded as a probabilistic theory over a non-distributive lattice~\cite{BirkhNeumLQM}, then it is \emph{not} the probabilistic trait that is problematic: 
The joint system~$F\otimes S$ is in a superposition, not a statistical mixture, as it might have also been the case in a probabilistic theory over a Boolean lattice.
If it were in a statistical mixture, there would not be a problem; it is the \emph{contextual} character of a non-distributive lattice that is essential to the measurement problem.\footnote{The aspect reflects in the importance of the joint system~$F\otimes S$ being isolated.
Decoherence ``dissolves'' the problem: One photon escaping from the friend's lab suffices for the ambiguity of the probability distribution to vanish.}
In this regard, it is secondary whether the problem becomes apparent in one or merely in a limit of multiple runs.
So after the extended Wigner's friend experiment, quantum mechanics is just as broken as it has been before---\emph{or not}:
\emph{If the terms ``isolated system,'' ``interaction,'' and ``measurement'' are used carefully, there is less of a problem}~\cite{HW18, HW19}.
The problem merely appears within a certain perspective---under certain unnecessary assumptions.

\emph{Contextuality} can be regarded as a consequence of subjecting interactions themselves to experience~\cite{HW19}, and, thus, rendering them meaningful~\cite{Nietzsche1873,WittgPhiloUntersuchungen,RortyCIS}.\footnote{Philosophy of language~\cite{Nietzsche1873,WittgPhiloUntersuchungen} has excavated a circular and intricate dependency between \emph{meaning} on the one side, and \emph{experience} on the other: A description of the world is meaningful insofar as it reflects experience. 
Conversely, any perceived feature of the world is inherent to a particular way of describing the world.
Perceiving and describing the world cannot be separated. 
Our perceptions are not encoded in an extra-discursive ``language of nature''~\cite{RortyCIS} but dependent on our ways of describing the world.}
Then, the measurement problem is not a mere insufficiency of quantum mechanics but rather inherent to particular requirements for physical theories---inherent to a specific way of picturing the world.
The measurement problem, thus, begs to reflect on the basic assumption and aspirations of doing physics.
 \section{The ``measurement'' problem}\label{sec:perspectives_on_the_observer}
\noindent
Maudlin's reading of the measurement problem~\cite{Maudlin95} is the inconsistency of the following triad: 
\begin{enumerate*}[label={(\alph*)}]
  \item\label{maudlin1} The time evolution of isolated systems in quantum mechanics is unitary;\footnote{The corresponding statement in~\cite{Maudlin95} refers to \emph{linear} time evolution following from the Schr\"{o}dinger equation. The inconsistency arises merely if Wigner's friend, measuring a system, is isolated, and, thus, the evolution modelled by a unitary operator;}
  \item\label{maudlin2} measurement results are \emph{exclusively one} of several possible values;
  \item\label{maudlin3} quantum mechanics is complete.
\end{enumerate*}
Statements~\ref{maudlin1} and~\ref{maudlin2} are fairly undebated: 
The first is a matter of definition\footnote{Collapse theories, such as GRW~\cite{GRW}, give rise to different statements and can, thus, be considered as formally different theories.}, and the latter a necessary requirement for falsifying theories~\cite{Popper1934}.
The last statement~\ref{maudlin3}, however, is a relict of the \emph{hegemonic aspirations} of physics rooted in a traditional understanding insinuating a correspondence of a theory's symbols with reality:
A~``measurement'' does here not only mean an interaction between two systems, but is the \emph{exhaustive description of an observation}---the account of experience that yields the normative empirical evidence for the validity of physical theories, including, paradoxically, quantum mechanics itself.
If, however, there is no privileged language, e.g., for reasons rooted in the nature of meaning and reference~\cite{WittgPhiloUntersuchungen,witt:trac22,RortyPMN,putnam1981,putnam1991representation}, then quantum mechanics does not have such a privileged status either:
The reduction of an observation (a meaningful account of experience) to symbols of a formal language is in doubt.
A measurement as an interaction between two systems is, in this regard, \emph{not equivalent} to an observation as an account of experience constitutive to meaning~\cite{HW18}:
\emph{The measurement is not so much of a problem, after all.}

 \section{Changing perspective}\label{sec:interaction_assumption}
\noindent
Whether or not there is an actual problem that needs to be solved, one may ask: 
What are the characteristics of theories that are troubled by the measurement problem?
Under which circumstances are we to expect a measurement problem?
What is so peculiar about quantum mechanics that it gives rise to the ``measurement problem''?
What is so peculiar about classical mechanics that it does \emph{not}?
If we cannot make sense of the measurement problem, then what is a perspective in which we \emph{can}?

If one requires 
\begin{enumerate*}[label={(\alph*)}]
  \item physical theories to account for interactions so that they are empirically significant, and 
  \item that an observation necessarily goes with such an interaction,
\end{enumerate*}
one can develop a language game that leads to probabilistic and contextual theories~\cite{HW19}. 
A ``Born rule'' that assigns probability weights to ordered sets of measurements as well as the measurement problem naturally accompany such theories.
Then, in fact, \emph{classical} mechanics seems strange as it is a theory without the ability to empirically trace the interaction---the causal link---necessary for sensory perceptions:
In classical mechanics, a system has determinate position and momentum if it is sufficiently characterized.
It is assumed that ``reading off'' the value of position and momentum does \emph{not} rely on an interaction---and potentially disturbance of the system---in a way that is captured by classical mechanics itself.\footnote{Dewey remarks that the quest for absolute and unchangeable truth drives towards a ``spectator theory:'' \blockcquote[{\S}1, p.26]{DeweyQFC}[.]{The common essence of all these theories, in short, is that what is known is antecedent to the mental act of observation and inquiry, and it totally unaffected by these acts; 
    otherwise it would not be fixed and unchangeable. [\ldots]
    The object refracts light to the eye and is seen; it makes a difference to the eye and to the person having an optical apparatus, but none to the thing seen.
    The real object is the object so fixed in its regal aloofness that it is a king to any beholding mind that may gaze upon it.
    A spectator theory of knowledge is the inevitable outcome}
   The interaction assumption might not be perspective easily embraced by physicists if it limits the ability to attain absolute knowledge.
    }
In fact, if it were, then the system's behavior would be indeterminate~\cite{PopperIndet1}. 
Thus, classical mechanics does not provide the ability to detect measurements.
In other words: There is no key-agreement protocol like ``BB84''~\cite{BB84}.
 \section{Knowledge --- Power}\label{sec:knowledge_power}
\noindent
Why is it appealing to embrace the last of Maudlin's statements, i.e., the assumption that quantum mechanics---\allowbreak or any other physical theory, for that matter---is complete?
What tempts to believe that quantum mechanics can provide an exhaustive description of an observer's account of experience---that quantum mechanics can excavate the ``ultimate real truth''?
If physics, and in particular quantum mechanics, is attributed a privileged access to reality, then physics attains a special and powerful role.
The unwillingness to even raise doubts about the exhaustive completeness of physical theories can be considered from a perspective of \emph{power and influence}:
Such doubts might undermine this special authority.
The link between discursive authority and a scientific language becomes apparent in Foucault's discourse analysis. 
For instance, the focus of a ``science of the subject'' has been narrowed down on \emph{sexuality}---\textcquote[p.\,70]{FoucaultTHOS}{[n]ot, however, by reason of some natural property inherent in sex itself, but by virtue of the tactics of power immanent in this discourse}.
The result: \textcquote[p.\,78]{FoucaultTHOS}[.]{We have placed ourselves under the sign of sex, but in the form of a \emph{Logic of Sex}, rather than a \emph{Physics}. [\ldots]
Whenever it is a question of knowing who we are, it is this logic that henceforth serves as our master key}
Sexuality could become the all-ruling language \emph{because} it attained the status of a science, with a ``logic'' of its own, and the ability to produce real and exhaustive knowledge.
This iterates the myth of a privileged access to reality and a derived privileged discursive status of science---a political extension to the myth of the given.
But it reveals at the same time that there is not \emph{one} scientific language we are inevitably converging to: 
Instead, it is a question of power tactics \emph{what} to put into the scientific focus---in this case, how to choose a perspective onto the subject.

Does ``putting ourselves under the sign of a Physics'' liberate us from the struggle for semantic authority?
Feyerabend~\cite{FeyerWDMZ} and Kuhn~\cite{kuhn1962structure} debunk the idea of physics being somewhat ``better'' than other scientific disciplines.
And why should it be?\footnote{If what we \emph{mean} is not an immaterial property---as observed by Rorty---, then we can also not conclude from some theory alone on the conclusions of an observer, as demanded by ``consistent reasoning'' in~\cite{FR18}.}
\blockcquote[{\S}1.2]{RortyPMN}[.]{It seems perfectly clear, at least since Wittgenstein and Sellars, that the `meaning' of typographical inscription is not an extra `immaterial' property they have, but just their place in a context of surrounding events in a language-game, in a form of life.
This goes for brain-inscriptions as well}
 
What are the consequences of embracing a linguistic ``master key'' to knowledge and truth about the subject?
A new master key is \emph{game theory} with its fundamental paradigm of the rational and egoistic player---the \emph{homo economicus}. 
To conclusively ``understand the subject,'' we merely have to know his objective function.
The view appeals also to physicists~\cite{Wallace2009}.
Schirrmacher\footnote{The epigraph of~\cite{SchirrmEgo} is, not surprisingly, a quote from Foucault: ``We should not try to discover who we are, but instead who we refuse to be.''~[own translation]} investigates~\cite{SchirrmEgo, FefeFrankSchirr} how modelling the subject as an egoist agent pursuing the maximum of his quantifiable objective is a discourse that \emph{produces egoists}---a discourse that unconditionally equates ``sensible'' with ``egoistic.''\footnote{\blockcquote[\S6]{SchirrmEgo}[.]{Das Problem ist, dass die Theorie nicht nur Handeln beschreibt, sondern Handeln erzwingt, sie ist nicht nur deskriptiv sondern normativ.
Sie postuliert nicht nur Egoisten, sie produziert sie.
Die Rationalit\"{a}t, die sie sich auf die Fahnen schreibt, kommt nicht von selbst. [\ldots]
Vern\"{u}nftiges Verhalten des Gegners entsteht nicht durch vern\"{u}nftige Argumente, sondern durch Drohung und Angst vor Vernichtung}
  In short: Fear-mongering aligns the agents with his game-theoretic model.
With the application of game theory to economics the agents have shifted from states to persons---and so have the targets of the alignment process.}
If we replace that key by artificial intelligence~\cite{patternDiscrim}: 
The problem remains. 
The way we look at things or subjects shapes what and how we see it.\footnote{Schirrmacher sees close links between the game-theoretic modelling of economical agents and a new ``information capitalism'' employing the tools of artificial intelligence.
Information is an advantage in the ``rationalized'' Game of Life.~\cite[\S6]{SchirrmEgo}}
\emph{The search for absolute truth has normalising, ordering, and narrowing effects.}

We have been referring to \emph{subjects}, complex human beings.
The described effects may not apply to the world of physical objects governed by the physical laws discovered by physics.\footnote{Already here there is an objection: How do we know that the observer's account of experience is part of that world of physical objects?}
On a fundamental level, we may still ``be'' quantum systems.
So are we then quantum brains-in-a-vat, connected by quantum channels? 
Is all we ever refer to---all we mean if we talk about things---eventually not an external thing, but a quantum state~\cite{putnam1981}?
This undermines what we essentially consider ``doing physics:''
\begin{enumerate}
  \item Doing physics \emph{establishes} meaning. It adds something to our language that is not inherent to it in the first place.
  \item The process of learning, understanding, and discovering continues. Quantum mechanics---and its current understanding---is not the end of physics. 
    We may expect---and search for---paradigm shifts as they happened before.
\end{enumerate}
Instead of trying to save quantum mechanics as the one privileged access to reality, or taking the measurement problem as an indicator that there must be another such privileged access, we advocate a pragmatic---in a philosophical sense---search for perspectives onto and descriptions of the world out there.
Any theory must have the modesty to allow for something in our experience that it cannot grasp.
No theory can claim for itself to have exhaustively captured an observer's account of experience, while it draws legitimacy from experimental findings.\footnote{We cannot escape the \emph{contingency of language}~\cite{RortyCIS}---we cannot step outside the various vocabularies: \emph{Also this text will remain a language game.}} 

\emph{Quantum mechanics is not the master key, but it is not broken either.}


\begin{thebibliography}{31}\makeatletter
\providecommand \@ifxundefined [1]{ \@ifx{#1\undefined}
}\providecommand \@ifnum [1]{ \ifnum #1\expandafter \@firstoftwo
 \else \expandafter \@secondoftwo
 \fi
}\providecommand \@ifx [1]{ \ifx #1\expandafter \@firstoftwo
 \else \expandafter \@secondoftwo
 \fi
}\providecommand \natexlab [1]{#1}\providecommand \enquote  [1]{``#1''}\providecommand \bibnamefont  [1]{#1}\providecommand \bibfnamefont [1]{#1}\providecommand \citenamefont [1]{#1}\providecommand \href@noop [0]{\@secondoftwo}\providecommand \href [0]{\begingroup \@sanitize@url \@href}\providecommand \@href[1]{\@@startlink{#1}\@@href}\providecommand \@@href[1]{\endgroup#1\@@endlink}\providecommand \@sanitize@url [0]{\catcode `\\12\catcode `\$12\catcode
  `\&12\catcode `\#12\catcode `\^12\catcode `\_12\catcode `\%12\relax}\providecommand \@@startlink[1]{}\providecommand \@@endlink[0]{}\providecommand \url  [0]{\begingroup\@sanitize@url \@url }\providecommand \@url [1]{\endgroup\@href {#1}{\urlprefix }}\providecommand \urlprefix  [0]{URL }\providecommand \Eprint [0]{\href }\providecommand \doibase [0]{http://dx.doi.org/}\providecommand \selectlanguage [0]{\@gobble}\providecommand \bibinfo  [0]{\@secondoftwo}\providecommand \bibfield  [0]{\@secondoftwo}\providecommand \translation [1]{[#1]}\providecommand \BibitemOpen [0]{}\providecommand \bibitemStop [0]{}\providecommand \bibitemNoStop [0]{.\EOS\space}\providecommand \EOS [0]{\spacefactor3000\relax}\providecommand \BibitemShut  [1]{\csname bibitem#1\endcsname}\let\auto@bib@innerbib\@empty
\bibitem [{\citenamefont {Bell}(1964)}]{Bell1964}  \BibitemOpen
  \bibfield  {author} {\bibinfo {author} {\bibfnamefont {John~S.}\ \bibnamefont
  {Bell}},\ }\bibfield  {title} {\enquote {\bibinfo {title} {{On the
  Einstein-Podolsky-Rosen paradox}},}\ }\href@noop {} {\bibfield  {journal}
  {\bibinfo  {journal} {Physics}\ }\textbf {\bibinfo {volume} {1}} (\bibinfo
  {year} {1964})}\BibitemShut {NoStop}\bibitem [{\citenamefont {Kochen}\ and\ \citenamefont {Specker}(1967)}]{KS67}  \BibitemOpen
  \bibfield  {author} {\bibinfo {author} {\bibfnamefont {Simon}\ \bibnamefont
  {Kochen}}\ and\ \bibinfo {author} {\bibfnamefont {Ernst}\ \bibnamefont
  {Specker}},\ }\bibfield  {title} {\enquote {\bibinfo {title} {The problem of
  hidden variables in quantum mechanics},}\ }\href@noop {} {\bibfield
  {journal} {\bibinfo  {journal} {Journal of Mathematics and Mechanics}\
  }\textbf {\bibinfo {volume} {17}},\ \bibinfo {pages} {59--87} (\bibinfo
  {year} {1967})}\BibitemShut {NoStop}\bibitem [{\citenamefont {Wigner}(1961)}]{Wigner1961}  \BibitemOpen
  \bibfield  {author} {\bibinfo {author} {\bibfnamefont {Eugene~P.}\
  \bibnamefont {Wigner}},\ }\bibfield  {title} {\enquote {\bibinfo {title}
  {Remarks on the mind-body question},}\ }in\ \href@noop {} {\emph {\bibinfo
  {booktitle} {The Scientist Speculates}}},\ \bibinfo {editor} {edited by\
  \bibinfo {editor} {\bibfnamefont {I.~J.}\ \bibnamefont {Good}}}\ (\bibinfo
  {publisher} {Heineman},\ \bibinfo {year} {1961})\BibitemShut {NoStop}\bibitem [{\citenamefont {Deutsch}(1985)}]{deutsch1985quantum}  \BibitemOpen
  \bibfield  {author} {\bibinfo {author} {\bibfnamefont {David}\ \bibnamefont
  {Deutsch}},\ }\bibfield  {title} {\enquote {\bibinfo {title} {Quantum theory
  as a universal physical theory},}\ }\href@noop {} {\bibfield  {journal}
  {\bibinfo  {journal} {International Journal of Theoretical Physics}\ }\textbf
  {\bibinfo {volume} {24}},\ \bibinfo {pages} {1--41} (\bibinfo {year}
  {1985})}\BibitemShut {NoStop}\bibitem [{\citenamefont {Frauchiger}\ and\ \citenamefont
  {Renner}(2018)}]{FR18}  \BibitemOpen
  \bibfield  {author} {\bibinfo {author} {\bibfnamefont {Daniela}\ \bibnamefont
  {Frauchiger}}\ and\ \bibinfo {author} {\bibfnamefont {Renato}\ \bibnamefont
  {Renner}},\ }\bibfield  {title} {\enquote {\bibinfo {title} {Quantum theory
  cannot consistently describe the use of itself},}\ }\href {\doibase
  10.1038/s41467-018-05739-8} {\bibfield  {journal} {\bibinfo  {journal}
  {Nature Communications}\ }\textbf {\bibinfo {volume} {9}} (\bibinfo {year}
  {2018}),\ 10.1038/s41467-018-05739-8}\BibitemShut {NoStop}\bibitem [{\citenamefont {Frauchiger}\ and\ \citenamefont
  {Renner}(2016)}]{FrRen}  \BibitemOpen
  \bibfield  {author} {\bibinfo {author} {\bibfnamefont {Daniela}\ \bibnamefont
  {Frauchiger}}\ and\ \bibinfo {author} {\bibfnamefont {Renato}\ \bibnamefont
  {Renner}},\ }\bibfield  {title} {\enquote {\bibinfo {title} {Single-world
  interpretations of quantum theory cannot be self-consistent},}\ }\href@noop
  {} {\  (\bibinfo {year} {2016})},\ \Eprint {http://arxiv.org/abs/1604.07422}
  {arXiv:1604.07422 [quant-ph]} \BibitemShut {NoStop}\bibitem [{\citenamefont {{Baumann}}\ \emph {et~al.}(2016)\citenamefont
  {{Baumann}}, \citenamefont {{Hansen}},\ and\ \citenamefont {{Wolf}}}]{BHW16}  \BibitemOpen
  \bibfield  {author} {\bibinfo {author} {\bibfnamefont {Veronika}\
  \bibnamefont {{Baumann}}}, \bibinfo {author} {\bibfnamefont {Arne}\
  \bibnamefont {{Hansen}}}, \ and\ \bibinfo {author} {\bibfnamefont {Stefan}\
  \bibnamefont {{Wolf}}},\ }\bibfield  {title} {\enquote {\bibinfo {title}
  {{The measurement problem is the measurement problem is the measurement
  problem}},}\ }\href@noop {} {\  (\bibinfo {year} {2016})},\ \Eprint
  {http://arxiv.org/abs/1611.01111} {arXiv:1611.01111 [quant-ph]} \BibitemShut
  {NoStop}\bibitem [{\citenamefont {Baumann}\ and\ \citenamefont {Wolf}(2018)}]{BW17}  \BibitemOpen
  \bibfield  {author} {\bibinfo {author} {\bibfnamefont {Veronika}\
  \bibnamefont {Baumann}}\ and\ \bibinfo {author} {\bibfnamefont {Stefan}\
  \bibnamefont {Wolf}},\ }\bibfield  {title} {\enquote {\bibinfo {title} {On
  formalisms and interpretations},}\ }\href {\doibase 10.22331/q-2018-10-15-99}
  {\bibfield  {journal} {\bibinfo  {journal} {{Quantum}}\ }\textbf {\bibinfo
  {volume} {2}},\ \bibinfo {pages} {99} (\bibinfo {year} {2018})}\BibitemShut
  {NoStop}\bibitem [{\citenamefont {Birkhoff}\ and\ \citenamefont {von
  Neumann}(1936)}]{BirkhNeumLQM}  \BibitemOpen
  \bibfield  {author} {\bibinfo {author} {\bibfnamefont {Garrett}\ \bibnamefont
  {Birkhoff}}\ and\ \bibinfo {author} {\bibfnamefont {John}\ \bibnamefont {von
  Neumann}},\ }\bibfield  {title} {\enquote {\bibinfo {title} {The logic of
  quantum mechanics},}\ }\href@noop {} {\bibfield  {journal} {\bibinfo
  {journal} {Annals of Mathematics}\ }\textbf {\bibinfo {volume} {37}},\
  \bibinfo {pages} {823--843} (\bibinfo {year} {1936})}\BibitemShut {NoStop}\bibitem [{\citenamefont {{Hansen}}\ and\ \citenamefont {{Wolf}}(2018)}]{HW18}  \BibitemOpen
  \bibfield  {author} {\bibinfo {author} {\bibfnamefont {Arne}\ \bibnamefont
  {{Hansen}}}\ and\ \bibinfo {author} {\bibfnamefont {Stefan}\ \bibnamefont
  {{Wolf}}},\ }\bibfield  {title} {\enquote {\bibinfo {title} {{The measurement
  problem is the `measurement' problem}},}\ }\href@noop {} {\  (\bibinfo {year}
  {2018})},\ \Eprint {http://arxiv.org/abs/1810.04573} {arXiv:1810.04573
  [quant-ph]} \BibitemShut {NoStop}\bibitem [{\citenamefont {{Hansen}}\ and\ \citenamefont {{Wolf}}(2019)}]{HW19}  \BibitemOpen
  \bibfield  {author} {\bibinfo {author} {\bibfnamefont {Arne}\ \bibnamefont
  {{Hansen}}}\ and\ \bibinfo {author} {\bibfnamefont {Stefan}\ \bibnamefont
  {{Wolf}}},\ }\bibfield  {title} {\enquote {\bibinfo {title} {Contextuality:
  Feature, not bug},}\ }\href@noop {} {\  (\bibinfo {year} {2019})}\BibitemShut
  {NoStop}\bibitem [{\citenamefont {Nietzsche}(1873)}]{Nietzsche1873}  \BibitemOpen
  \bibfield  {author} {\bibinfo {author} {\bibfnamefont {Friedrich~Wilhelm}\
  \bibnamefont {Nietzsche}},\ }\href@noop {} {\emph {\bibinfo {title} {{\"U}ber
  Wahrheit und L{\"u}ge im aussermoralischen Sinne}}}\ (\bibinfo {year}
  {1873})\BibitemShut {NoStop}\bibitem [{\citenamefont {Wittgenstein}(1953)}]{WittgPhiloUntersuchungen}  \BibitemOpen
  \bibfield  {author} {\bibinfo {author} {\bibfnamefont {Ludwig}\ \bibnamefont
  {Wittgenstein}},\ }\href@noop {} {\emph {\bibinfo {title} {Philosophische
  Untersuchungen}}},\ edited by\ \bibinfo {editor} {\bibfnamefont {Joachim}\
  \bibnamefont {Schulte}}\ (\bibinfo  {publisher} {Suhrkamp Verlag},\ \bibinfo
  {year} {1953})\BibitemShut {NoStop}\bibitem [{\citenamefont {Rorty}(1989)}]{RortyCIS}  \BibitemOpen
  \bibfield  {author} {\bibinfo {author} {\bibfnamefont {Richard}\ \bibnamefont
  {Rorty}},\ }\href@noop {} {\emph {\bibinfo {title} {Contingency, Irony, and
  Solidarity}}}\ (\bibinfo  {publisher} {Cambridge University Press},\ \bibinfo
  {year} {1989})\BibitemShut {NoStop}\bibitem [{\citenamefont {Maudlin}(1995)}]{Maudlin95}  \BibitemOpen
  \bibfield  {author} {\bibinfo {author} {\bibfnamefont {Tim}\ \bibnamefont
  {Maudlin}},\ }\bibfield  {title} {\enquote {\bibinfo {title} {Three
  measurement problems},}\ }\href {\doibase 10.1007/BF00763473} {\bibfield
  {journal} {\bibinfo  {journal} {Topoi}\ }\textbf {\bibinfo {volume} {14}},\
  \bibinfo {pages} {7--15} (\bibinfo {year} {1995})}\BibitemShut {NoStop}\bibitem [{\citenamefont {Ghirardi}\ \emph {et~al.}(1985)\citenamefont
  {Ghirardi}, \citenamefont {Rimini},\ and\ \citenamefont {Weber}}]{GRW}  \BibitemOpen
  \bibfield  {author} {\bibinfo {author} {\bibfnamefont {Giancarlo}\
  \bibnamefont {Ghirardi}}, \bibinfo {author} {\bibfnamefont {Alberto}\
  \bibnamefont {Rimini}}, \ and\ \bibinfo {author} {\bibfnamefont {Tullio}\
  \bibnamefont {Weber}},\ }\bibfield  {title} {\enquote {\bibinfo {title} {A
  model for a unified quantum description of macroscopic and microscopic
  systems},}\ }in\ \href@noop {} {\emph {\bibinfo {booktitle} {Quantum
  Probability and Applications II}}},\ \bibinfo {editor} {edited by\ \bibinfo
  {editor} {\bibfnamefont {Luigi}\ \bibnamefont {Accardi}}\ and\ \bibinfo
  {editor} {\bibfnamefont {Wilhelm}\ \bibnamefont {von Waldenfels}}}\ (\bibinfo
   {publisher} {Springer},\ \bibinfo {address} {Berlin, Heidelberg},\ \bibinfo
  {year} {1985})\ pp.\ \bibinfo {pages} {223--232}\BibitemShut {NoStop}\bibitem [{\citenamefont {Popper}(1934)}]{Popper1934}  \BibitemOpen
  \bibfield  {author} {\bibinfo {author} {\bibfnamefont {Karl}\ \bibnamefont
  {Popper}},\ }\href@noop {} {\emph {\bibinfo {title} {Logik der Forschung}}},\
  \bibinfo {edition} {11th}\ ed.\ (\bibinfo  {publisher} {Mohr Siebeck},\
  \bibinfo {address} {T{\"u}bingen},\ \bibinfo {year} {1934})\BibitemShut
  {NoStop}\bibitem [{\citenamefont {Wittgenstein}(1922)}]{witt:trac22}  \BibitemOpen
  \bibfield  {author} {\bibinfo {author} {\bibfnamefont {Ludwig}\ \bibnamefont
  {Wittgenstein}},\ }\href@noop {} {\emph {\bibinfo {title} {{Tractatus
  logico-philosophicus}}}}\ (\bibinfo  {publisher} {Routledge},\ \bibinfo
  {year} {1922})\BibitemShut {NoStop}\bibitem [{\citenamefont {Rorty}(1979)}]{RortyPMN}  \BibitemOpen
  \bibfield  {author} {\bibinfo {author} {\bibfnamefont {Richard}\ \bibnamefont
  {Rorty}},\ }\href@noop {} {\emph {\bibinfo {title} {Philosophy and the Mirror
  of Nature}}}\ (\bibinfo  {publisher} {Princeton University Press},\ \bibinfo
  {year} {1979})\BibitemShut {NoStop}\bibitem [{\citenamefont {Putnam}(1981)}]{putnam1981}  \BibitemOpen
  \bibfield  {author} {\bibinfo {author} {\bibfnamefont {Hilary}\ \bibnamefont
  {Putnam}},\ }\href@noop {} {\emph {\bibinfo {title} {Reason, Truth and
  History}}},\ Philosophical Papers\ (\bibinfo  {publisher} {Cambridge
  University Press},\ \bibinfo {year} {1981})\BibitemShut {NoStop}\bibitem [{\citenamefont {Putnam}(1991)}]{putnam1991representation}  \BibitemOpen
  \bibfield  {author} {\bibinfo {author} {\bibfnamefont {Hilary}\ \bibnamefont
  {Putnam}},\ }\href@noop {} {\emph {\bibinfo {title} {Representation and
  Reality}}},\ A Bradford book\ (\bibinfo  {publisher} {A Bradford Book},\
  \bibinfo {year} {1991})\BibitemShut {NoStop}\bibitem [{\citenamefont {Dewey}(1929)}]{DeweyQFC}  \BibitemOpen
  \bibfield  {author} {\bibinfo {author} {\bibfnamefont {John}\ \bibnamefont
  {Dewey}},\ }\href@noop {} {\emph {\bibinfo {title} {The Quest for Certainty:
  A Study of the Relation of Knowledge and Action}}}\ (\bibinfo  {publisher}
  {Minton, Balch and Company},\ \bibinfo {year} {1929})\BibitemShut {NoStop}\bibitem [{\citenamefont {Popper}(1950)}]{PopperIndet1}  \BibitemOpen
  \bibfield  {author} {\bibinfo {author} {\bibfnamefont {Karl}\ \bibnamefont
  {Popper}},\ }\bibfield  {title} {\enquote {\bibinfo {title} {Indeterminism in
  quantum physics and in classical physics 1},}\ }\href {\doibase
  10.1093/bjps/I.2.117} {\bibfield  {journal} {\bibinfo  {journal} {Brit.
  Journ. for the Phil. of Sci.}\ }\textbf {\bibinfo {volume} {I}},\ \bibinfo
  {pages} {117--133} (\bibinfo {year} {1950})}\BibitemShut {NoStop}\bibitem [{\citenamefont {Bennett}\ and\ \citenamefont
  {Brassard}(1984)}]{BB84}  \BibitemOpen
  \bibfield  {author} {\bibinfo {author} {\bibfnamefont {Charles}\ \bibnamefont
  {Bennett}}\ and\ \bibinfo {author} {\bibfnamefont {Gilles}\ \bibnamefont
  {Brassard}},\ }\bibfield  {title} {\enquote {\bibinfo {title} {{Quantum
  cryptography: Public key distribution and coin tossing}},}\ }in\ \href@noop
  {} {\emph {\bibinfo {booktitle} {Proceedings of IEEE International Conference
  on Computers, Systems, and Signal Processing}}}\ (\bibinfo {year} {1984})\
  p.\ \bibinfo {pages} {175}\BibitemShut {NoStop}\bibitem [{\citenamefont {Foucault}(1978)}]{FoucaultTHOS}  \BibitemOpen
  \bibfield  {author} {\bibinfo {author} {\bibfnamefont {Michel}\ \bibnamefont
  {Foucault}},\ }\href@noop {} {\emph {\bibinfo {title} {The History of
  Sexuality: An Introduction}}},\ Histoire de la sexualit{\'e}\ (\bibinfo
  {publisher} {Pantheon Books},\ \bibinfo {year} {1978})\BibitemShut {NoStop}\bibitem [{\citenamefont {Feyerabend}(1986)}]{FeyerWDMZ}  \BibitemOpen
  \bibfield  {author} {\bibinfo {author} {\bibfnamefont {Paul}\ \bibnamefont
  {Feyerabend}},\ }\href@noop {} {\emph {\bibinfo {title} {{Wider den
  Methodenzwang}}}}\ (\bibinfo  {publisher} {Suhrkamp Verlag},\ \bibinfo {year}
  {1986})\BibitemShut {NoStop}\bibitem [{\citenamefont {Kuhn}(1962)}]{kuhn1962structure}  \BibitemOpen
  \bibfield  {author} {\bibinfo {author} {\bibfnamefont {Thomas~Samuel}\
  \bibnamefont {Kuhn}},\ }\href {https://books.google.ch/books?id=zsfMvwEACAAJ}
  {\emph {\bibinfo {title} {The Structure of Scientific Revolutions}}},\
  International encyclopedia of unified science\ (\bibinfo  {publisher}
  {University of Chicago Press},\ \bibinfo {year} {1962})\BibitemShut {NoStop}\bibitem [{\citenamefont {{Wallace}}(2009)}]{Wallace2009}  \BibitemOpen
  \bibfield  {author} {\bibinfo {author} {\bibfnamefont {David}\ \bibnamefont
  {{Wallace}}},\ }\bibfield  {title} {\enquote {\bibinfo {title} {{A formal
  proof of the Born rule from decision-theoretic assumptions}},}\ }\href@noop
  {} {\  (\bibinfo {year} {2009})},\ \Eprint {http://arxiv.org/abs/0906.2718}
  {arXiv:0906.2718 [quant-ph]} \BibitemShut {NoStop}\bibitem [{\citenamefont {Schirrmacher}(2013)}]{SchirrmEgo}  \BibitemOpen
  \bibfield  {author} {\bibinfo {author} {\bibfnamefont {Frank}\ \bibnamefont
  {Schirrmacher}},\ }\href@noop {} {\emph {\bibinfo {title} {Ego: das Spiel des
  Lebens}}}\ (\bibinfo  {publisher} {Karl Blessing Verlag},\ \bibinfo {year}
  {2013})\BibitemShut {NoStop}\bibitem [{\citenamefont {von Leitner}\ \emph {et~al.}(2013)\citenamefont {von
  Leitner}, \citenamefont {Schirrmacher},\ and\ \citenamefont
  {Rieger}}]{FefeFrankSchirr}  \BibitemOpen
  \bibfield  {author} {\bibinfo {author} {\bibfnamefont {Felix}\ \bibnamefont
  {von Leitner}}, \bibinfo {author} {\bibfnamefont {Frank}\ \bibnamefont
  {Schirrmacher}}, \ and\ \bibinfo {author} {\bibfnamefont {Frank}\
  \bibnamefont {Rieger}},\ }\href {https://alternativlos.org/29/} {\enquote
  {\bibinfo {title} {{Alternativlos 35}},}\ }\bibinfo {howpublished}
  {\url{https://alternativlos.org/29/}} (\bibinfo {year} {2013}),\ \bibinfo
  {note} {[Online; accessed 20-Sep-2018]}\BibitemShut {NoStop}\bibitem [{\citenamefont {Apprich}\ \emph {et~al.}(2018)\citenamefont
  {Apprich}, \citenamefont {Cramer}, \citenamefont {Kyong~Chun},\ and\
  \citenamefont {Steyerl}}]{patternDiscrim}  \BibitemOpen
  \bibfield  {author} {\bibinfo {author} {\bibfnamefont {Clemens}\ \bibnamefont
  {Apprich}}, \bibinfo {author} {\bibfnamefont {Florian}\ \bibnamefont
  {Cramer}}, \bibinfo {author} {\bibfnamefont {Wendy~Hui}\ \bibnamefont
  {Kyong~Chun}}, \ and\ \bibinfo {author} {\bibfnamefont {Hito}\ \bibnamefont
  {Steyerl}},\ }\href@noop {} {\emph {\bibinfo {title} {Pattern
  Discrimination}}}\ (\bibinfo  {publisher} {meson press},\ \bibinfo {year}
  {2018})\BibitemShut {NoStop}\end{thebibliography}
\end{document}